\documentclass[galaxies,article,accept,moreauthors,pdftex,10pt,a4paper]{mdpi} 

\usepackage{amsmath}
\usepackage{amssymb}

\firstpage{1} 
\articlenumber{x}
\doinum{10.3390/------}
\pubvolume{xx}
\pubyear{2017}
\copyrightyear{2017}
\externaleditor{Academic Editor: ---}
\history{Received: 14 September 2017; Accepted: 06 October 2017; Published: ---}

\Title{RADIO POLARISATION STUDY OF HIGH ROTATION MEASURE AGNS}

\orcidauthorONE{0000-0003-0742-2006} % Add \orcidONE{} behind the author's name
\orcidauthorTWO{0000-0001-8906-7866} % Add \orcidTWO{} behind the author's name
\orcidauthorFIVE{0000-0001-7722-8458}

\Author{Yik Ki Ma $^{1,}$* \orcidONE{}, Sui Ann Mao $^{1}$ \orcidTWO{}, Aritra Basu $^{1}$, Carl Heiles $^{2}$ and Jennifer West $^{3}$ \orcidFIVE{} }

\AuthorNames{Yik Ki Ma, Sui Ann Mao, Aritra Basu, Carl Heiles and Jennifer West}

\address{%
$^{1}$ \quad Max-Planck-Institut f\"ur Radioastronomie, Auf dem H\"ugel 69, 53121 Bonn, Germany\\
$^{2}$ \quad Department of Astronomy, University of California, Berkeley, CA 94720-3411, USA\\
$^{3}$ \quad Dunlap Institute for Astronomy and Astrophysics, The University of Toronto, 50 St.\ George Street, Toronto, ON M5S 3H4, Canada}

\corres{Correspondence: ykma@mpifr-bonn.mpg.de; Tel.: +49-228-525-180}

\abstract{
As radio polarised emission from astrophysical objects traverse through foreground magnetised plasma, the physical conditions along the lines of sight are encrypted in the form of Rotation Measure (RM). We performed broadband spectro-polarimetric observations of high Rotation Measure ($|{\rm RM}| \gtrsim 300\,{\rm rad\,m}^{-2}$) sources away from the Galactic plane ($|b| > 10^\circ$) selected from the NVSS RM catalogue \citep{taylor09}. The main goals are to verify the NVSS RM values, which could be susceptible to $n\pi$-ambiguity, as well as to identify the origin of the extreme RM values. We show that $40\,\%$ of our sample suffer from $n\pi$-ambiguity in the NVSS RM catalogue. There are also hints of RM variabilities over $\sim 20$ years epoch for most of our sources, as revealed by comparing the RM values of the two studies in the same frequency ranges after correcting for $n\pi$-ambiguity. At last, we demonstrate the possibility of applying \textit{QU}-fitting to study the ambient media of AGNs.}

\keyword{galaxies: active; galaxies: magnetic fields; ISM: magnetic fields; magnetic fields; radio continuum: galaxies; techniques: polarimetric}

\begin{document}
%%%%%%%%%%%%%%%%%%%%%%%%%%%%%%%%%%%%%%%%%%
%% Only for the journal Gels: Please place the Experimental Section after the Conclusions

%%%%%%%%%%%%%%%%%%%%%%%%%%%%%%%%%%%%%%%%%%
% \setcounter{section}{-1} %% Remove this when starting to work on the template.
% \section{How to Use this Template}
% The template details the sections that can be used in a manuscript. Note that the order of article sections may differ from the requirements of the journal (e.g. for the positioning of the Materials and Methods section). Please check the instructions for authors page of the journal to verify the correct order. For any questions, please contact the editorial office of the journal or support@mdpi.com. For LaTeX related questions please contact Janine Daum at latex-support@mdpi.com.

\section{Introduction}

Cosmic magnetic fields are important in numerous astrophysical processes. For example, star formation rates in galaxies are regulated by magnetic pressures. The propagation of cosmic ray particles are constrained by magnetic fields in galaxies. Magnetic fields can also promote outflows of hot gas from galaxies by providing the necessary pressure. However, fundamental questions of cosmic magnetism such as the large-scale magnetic field geometry in the disk of the Milky Way are not completely understood \citep[see e.g.][]{beck16}.

One way to characterise cosmic magnetic fields is to use the Faraday rotation experienced by emission from polarised sources. The emission is said to be originating from a certain Faraday depth (FD [${\rm rad\,m}^{-2}$]) if it experiences a rotation of the polarisation angle (PA [rad]) of:

\begin{equation}
\Delta {\rm PA} = \left[ 0.81 \int_\ell^0 n_e(s) B_{||}(s)\,{\rm d}s \right] \cdot \lambda^2 \equiv {\rm FD} \cdot \lambda^2{\rm ,}
\end{equation}
where $\ell$ [pc] is the distance of the emission region from the observer, $n_e$ [cm$^{-3}$] is the electron density, $B_{||}$ [$\mu$G] is the magnetic field strength along the line of sight ($s$ [pc]), and $\lambda$ [m] is the observed wavelength. The physical conditions in the foreground magneto-ionised media, in particular $n_e$ and $B_{||}$, can thus be revealed by measuring the FD values of background polarised sources.

For the case where the polarised emission from a source originates from a single FD, the source is ``Faraday simple''. Sometimes, the sources are implicitly assumed to be Faraday simple because of observational constraints such as limited coverages in $\lambda^2$ space. In both situations, FD is equivalent to Rotation Measure (RM), which is the slope of PA against $\lambda^2$. The traditional way to obtain RM values of astrophysical objects is to measure PAs at two or more different wavelengths, and perform a linear fit to PA as a function of $\lambda^2$. For the simplest case of PA measurements at only two bands (denoted by subscript 1 and 2), the RM is given by

\begin{equation}
{\rm RM} = \frac{{\rm PA}_1 - {\rm PA}_2 + n\pi}{\lambda_1^2 - \lambda_2^2}{\rm ,}
\end{equation}
where $n$ is an integer corresponding to the $n\pi$-ambiguity. It arises because the PAs measured by our instruments can only fall within $\displaystyle[-\frac{\pi}{2}, \frac{\pi}{2}]$. This restricts the relative Faraday rotation (${\rm PA}_1 - {\rm PA}_2$) to $\displaystyle[-\pi, \pi]$, although in reality it can fall outside of this range. The $n\pi$ term removes this limitation but also introduces an ambiguity in the determination of RMs. In general, this ambiguity can be eliminated by having more PA measurements at different wavelengths. The broadband capabilities of modern radio interferometers [e.g.\ the Karl G.\ Jansky Very Large Array (VLA) and the Australia Telescope Compact Array (ATCA)] allow spectro-polarimetric observations with a continuous frequency coverage over a wide bandwidth (e.g.\ 1--2\,GHz in L-band and 2--4\,GHz in S-band for the VLA), which allows determination of FDs without ambiguities \citep[e.g.][]{brentjens05}.

The NVSS RM catalogue \cite{taylor09} contains RM values of 37,543 polarised radio sources at about 1400\,MHz. It covers the sky at $\delta > -40^\circ$ and serves as an invaluable tool for the study of Galactic magnetism --- including studies of the magnetic fields in Galactic H~{\sc ii} regions \citep[e.g.][]{harvey-smith11,purcell15} and the Galactic halo \citep[e.g.][]{terral17}. The catalogue is based on the original NVSS catalogue \citep{condon98}, with the survey conducted in two bands centred at 1364.9\,MHz and 1435.1\,MHz with bandwidths of 42\,MHz each. This leaves open the possibility of $n\pi$-ambiguity affecting some of the NVSS RM sources. An algorithm was adopted by Taylor et al.\ to minimise the effect of $n\pi$-ambiguity in the RM catalogue, by considering (1) the observed Faraday bandwidth depolarisation of each source, and (2) the median RM values of the surrounding sources. However, the effectiveness of this algorithm is not clear. One way to test this is to observe a small fraction of the NVSS RM sources again and compare the true FD values from new broadband observations with the corresponding RM values listed in the NVSS RM catalogue.

Here, we report the results from our broadband observations of 20 NVSS sources with the VLA. These sources are believed to be prone to the $n\pi$-ambiguity. The FD values obtained from our new observations are directly compared to the RMs in the NVSS RM catalogue. The observations and data reduction procedures are outlined in Section~\ref{sec:obs}. We present the preliminary results in Section~\ref{sec:res}. In Section~\ref{sec:dis}, we discuss the implications of our study, and demonstrate the potential of using the broadband data to study the ambient media surrounding our target sources, presumably radio Active Galactic Nuclei (AGNs). The study is summarised in Section~\ref{sec:con}.

 %%%%%%%%%%%%%%%%%%%%%%%%%%%%%%%%%%%%%%%%%%
\section{Observations and Data Reduction \label{sec:obs}}
 
\subsection{Source Selection}
We observed a sample of 20 sources from the NVSS RM catalogue \citep{taylor09}, with all of them situated off the Galactic plane ($|b| > 10^\circ$). They are considered as the prime candidates of $n\pi$-ambiguity in the catalogue, with most of them having high reported RM magnitudes ($\gtrsim 300\,{\rm rad\,m}^{-2}$). For comparison, the majority ($\sim 99\,\%$) of sources in the same region have much lower $|$RM$|$s ($< 150\,{\rm rad\,m}^{-2}$). It is therefore possible that their high $|$RM$|$ values simply stem from $n\pi$-ambiguity.

\subsection{Observational Setup and Data Reduction}
The observations were performed with the VLA in L-band (1--2\,GHz) with D array configuration on 03 July 2014. We adopted the standard snapshot polarisation observation strategy for data calibration. The on-source time was about 3--4 minutes per source. The Common Astronomy Software Applications (CASA) package \citep{mcmullin07} was used for all the data reduction procedures. Radio image cubes in Stokes \textit{I}, \textit{Q}, and \textit{U} were formed at a 4\,MHz interval in observed frequency, giving a typical rms noise of about $1\,{\rm mJy\,beam}^{-1}$ in Stokes \textit{I} and $0.5\,{\rm mJy\,beam}^{-1}$ in Stokes \textit{Q} and \textit{U}. From these data cubes, we extracted the flux densities in Stokes \textit{I}, \textit{Q}, and \textit{U} of our target sources for further analysis.
 
%%%%%%%%%%%%%%%%%%%%%%%%%%%%%%%%%%%%%%%%%%
\section{Results \label{sec:res}}

\subsection{RM-Synthesis \label{sec:rmsyn}}
We applied the RM-Synthesis technique \citep{brentjens05} to obtain the FD values of our target sources. \textit{Q/I} and \textit{U/I} values as a function of $\lambda^2$ are used to construct the Faraday spectra (which is polarisation fraction as a function of FD). A Python-based RM-Synthesis code\footnote{Available on \url{http://www.github.com/mrbell/pyrmsynth}.} was used to perform this analysis. The resulting Faraday spectra were deconvolved with the RM-Clean algorithm \citep[e.g.][]{heald09}, and then the top seven points of each spectrum are fitted with a parabola to extract the FD of the respective source \citep[e.g.][]{heald09,mao10}. This means we are taking the FD of the Faraday component with highest polarisation fraction as the source's FD. Nonetheless, we argue that the FD values obtained this way can still be reasonably compared with the NVSS RM values, and we will demonstrate the ability to apply \textit{QU}-fitting \citep[e.g.][]{osullivan12} to our data to unveil the Faraday complexities \citep[e.g.][]{osullivan12,anderson16} of our sources in Section~\ref{sec:qufit}.

\begin{figure}[H]
\centering
\includegraphics[width=0.49\textwidth]{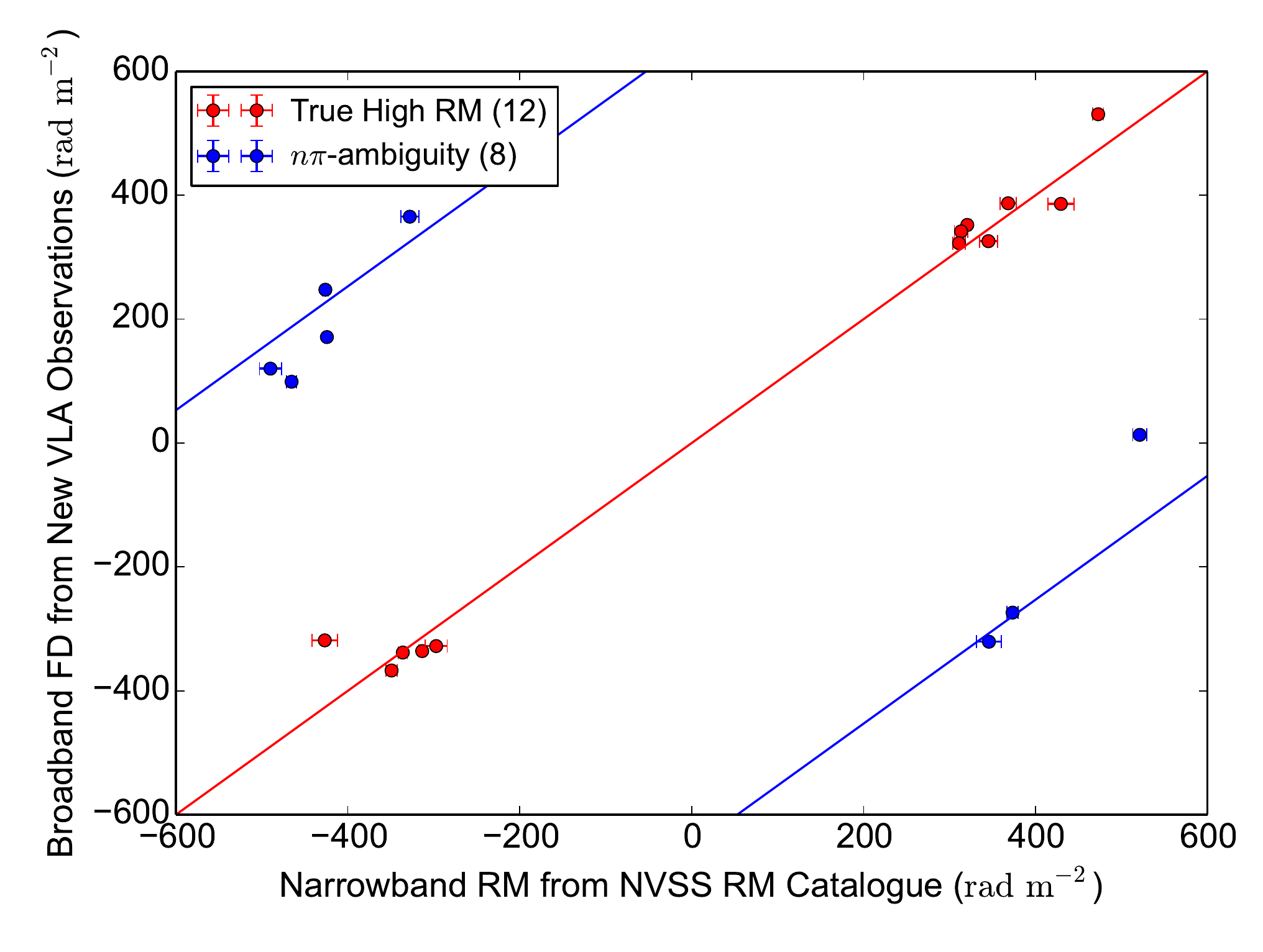}
\includegraphics[width=0.49\textwidth]{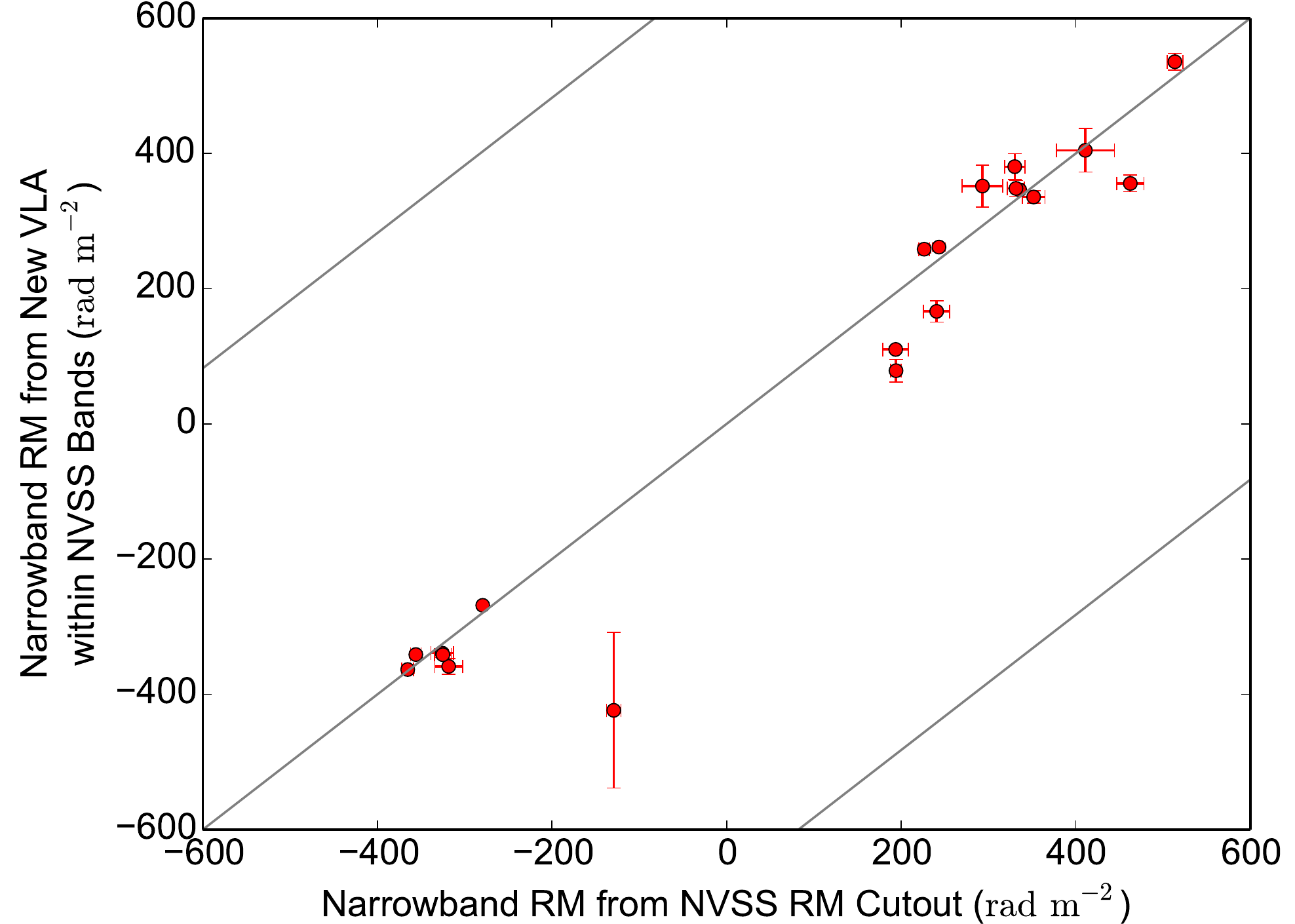}
\caption{\textbf{(Left)} FDs from new VLA observations plotted against the RM values in the NVSS RM catalogue. The data points should lie near the red line if FD/RM values from the two studies agree with each other (i.e.\ they are truly high RM sources), or one of the blue lines if they suffer from $1\pi$-ambiguity in the NVSS RM catalogue. \textbf{(Right)} RM from our new VLA observations using only the NVSS frequency bands against that obtained from NVSS RM cutout images. Both sets of RM values are obtained from performing the traditional $\lambda^2$ fit, with $n\pi$-ambiguity resolved using the broadband FD values. \label{fig:f1}}
\end{figure}

The FD values obtained from new VLA observations are compared with the RM values from the NVSS RM catalogue [Figure~\ref{fig:f1} (Left)]. We find that 12 sources have FD/RM values which agree with each other in the two measurements within about $\pm 100\,{\rm rad\,m}^{-2}$, while eight of them have differences in FD/RM by about $\pm 652.9\,{\rm rad\,m}^{-2}$. Since the frequency coverage of the two observations are vastly different, Faraday complexities can cause deviations of PA from the simple linear relationship with $\lambda^2$, and for such cases the FD/RM values determined can vary depending on the frequency coverages \citep[e.g.][]{anderson16}. More care is needed to facilitate a direct comparison between our broadband data and the narrowband NVSS data.

\subsection{Direct Comparison with NVSS RM Catalogue}

We formed a separate set of images from our new VLA data using only the frequency coverage of the NVSS survey. The RM values from the two observations are then derived from the images using identical methods, with the $n\pi$-ambiguity removed using the broadband FD values we obtained in Section~\ref{sec:rmsyn}. The comparison of the RM values between the two observations in the same frequency ranges are shown in Figure~\ref{fig:f1} (Right). The differences in RM of our sources have a mean of $2.4\,\sigma$, going up to $6.5\,\sigma$, with most ($85\,\%$) of them having RM differences of more than $1\,\sigma$, where $\sigma$ is the RM uncertainty of the two studies added in quadrature.

\subsection{QU-fitting \label{sec:qufit}}
The technique of \textit{QU}-fitting \citep[e.g.][]{osullivan12} attempts to capture the behaviour of Stokes \textit{Q} and \textit{U} as a function of $\lambda^2$ by fitting them with different Faraday rotation models. For our study we tried models including single/double Faraday simple, single/double Burn slab(s), and single Burn slab with foreground Faraday screen \citep{burn66,sokoloff98}.

\begin{figure}[H]
\centering
\includegraphics[width=0.99\textwidth]{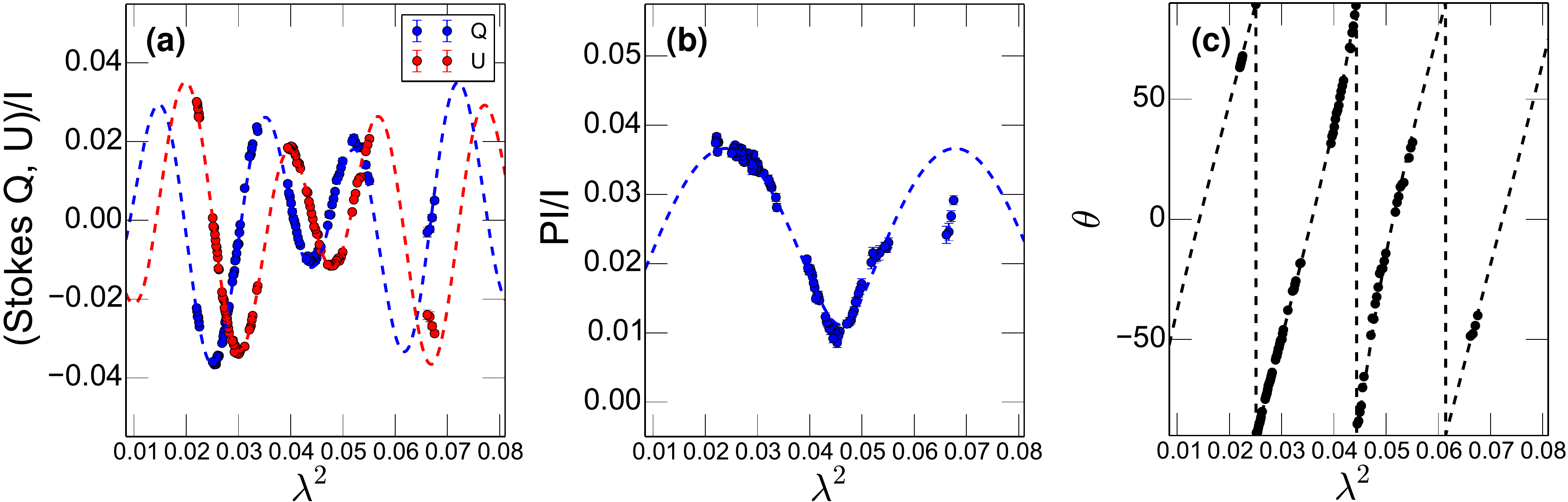}
\caption{\textit{QU}-fitting results of NVSS J190255+315942, with \textbf{(a)} \textit{Q}/\textit{I} and \textit{U}/\textit{I} against $\lambda^2$, \textbf{(b)} polarisation fraction against $\lambda^2$, and \textbf{(c)} PA against $\lambda^2$. The dotted lines represent the best-fit model for the source (see text). \label{fig:f3}}
\end{figure} 

In Figure~\ref{fig:f3}, we show our preliminary \textit{QU}-fitting results of NVSS J190255+315942 to demonstrate the capability of broadband spectro-polarimetry. The best-fit model, judged by the reduced $\chi^2$ value, consists of two Faraday simple components, with polarisation fractions of $2.39 \pm 0.02\,\%$ and $1.27 \pm 0.02\,\%$, and FDs of $+168.0 \pm 0.5\,{\rm rad\,m}^{-2}$ and $+96.5 \pm 0.9\,{\rm rad\,m}^{-2}$ respectively.
%%%%%%%%%%%%%%%%%%%%%%%%%%%%%%%%%%%%%%%%%%
\section{Discussion \label{sec:dis}}

\subsection{$n\pi$-ambiguity in the NVSS RM Catalogue}
We showed in Section~\ref{sec:rmsyn} that some of the sources have erroneous values in the RM catalogue \citep{taylor09} due to $n\pi$-ambiguity. It is unclear how many out of the total of 37,543 RMs may have the same issue, as we have only observed a small sample of sources which we considered likely to suffer from $n\pi$-ambiguity. The algorithm used for constructing the NVSS RM catalogue is likely to become less effective within $|$RM$|$ ranges of $< 50\,{\rm rad\,m}^{-2}$, $> 520\,{\rm rad\,m}^{-2}$, and near $326.5\,{\rm rad\,m}^{-2}$ \citep{taylor09}. Another possible reason of erroneous RM values is Faraday complexity \citep[e.g.][]{oppermann12}, as it can cause the PA to deviate from the linear relationship with $\lambda^2$. Future broadband radio polarisation surveys, such as the Very Large Array Sky Survey (VLASS) in S-band, could verify all the NVSS RM values.

\subsection{Possible RM Variabilities}
By comparing the narrowband RM values computed using our new VLA data to that from the NVSS RM images within the same frequency ranges, we found that the RM values of most sources do not agree within the measurement uncertainties between the two epochs. Such discrepancies could stem from systematic errors yet to be accounted for. For instance, while all the targets are placed at the pointing centres in our observations, NVSS data were taken with multi-pointing mosaic observations with our target sources not necessarily placed on-axis. This can result in significant polarisation leakage that is not easy to characterise \citep[e.g.][]{jagannathan17}. The polarisation leakage component can potentially change the observed PAs, causing the NVSS RM values to deviate from the true values. The RM differences between the two studies could also be attributed to true RM variabilities over the two epochs. If this is indeed the case, higher cadence campaigns of these sources can reveal the physical characteristics of the magneto-ionised media giving rise to such variabilities. 

\subsection{QU-fitting as Probe of Vicinities of AGN Jets}
We showed the \textit{QU}-fitting results of one of our target sources in Section~\ref{sec:qufit}. Despite it is not spatially resolved in our radio images (synthesised beam $\sim 45^{\prime\prime}$), we could still ``resolve'' it into two Faraday components by broadband spectro-polarimetric observations and \textit{QU}-fitting. One way to apply this result is to the study of the ambient media of the target source, presumably a radio-loud AGN. NVSS J190255+315942 can be resolved into two Faraday simple components with a difference of FD of $71.6 \pm 1.0\,{\rm rad\,m}^{-2}$ between them. Such a large difference in FD is not expected from the Milky Way within $45^{\prime\prime}$ \citep[e.g.][]{haverkorn08,mao10}, especially off the Galactic plane. One possibility is to attribute the two Faraday components to two radio lobes, with the jet axis tilted with respect to the plane of sky. In this picture, emission from the further lobe has to travel through extra magneto-ionised material surrounding the AGN, analogous to the Laing-Harrington effect \citep{laing88,garrington88}, and the difference in FD of the two components could originate from this extra material.
%%%%%%%%%%%%%%%%%%%%%%%%%%%%%%%%%%%%%%%%%%
\section{Summary \label{sec:con}}

In this proceedings, we presented our study of high $|$RM$|$ sources from the NVSS RM catalogue with new broadband spectro-polarimetric data. We found that $40\,\%$ of our targets have erroneous RM values in the RM catalogue. This can potentially impact studies of cosmic magnetism which utilises the catalogue. There are also hints of RM time variabilities in most of our sources, which can be verified by future broadband polarisation monitoring. At last, we demonstrated the possibility of applying the \textit{QU}-fitting technique on broadband spectro-polarimetric data to study the ambient media of AGNs.

%%%%%%%%%%%%%%%%%%%%%%%%%%%%%%%%%%%%%%%%%%
\vspace{6pt} 

%%%%%%%%%%%%%%%%%%%%%%%%%%%%%%%%%%%%%%%%%%
%% optional
%\supplementary{The following are available online at www.mdpi.com/link, Figure S1: title, Table S1: title, Video S1: title.}

%%%%%%%%%%%%%%%%%%%%%%%%%%%%%%%%%%%%%%%%%%
\acknowledgments{We thank Jeroen Stil for providing us with the NVSS RM cutout images. We thank Rainer Beck, Shane O'Sullivan, and Jeroen Stil for helpful discussion on this work. YKM was supported for this research by the International Max Planck Research School (IMPRS) for Astronomy and Astrophysics at the University of Bonn and Cologne. YKM acknowledges partial support through the Bonn-Cologne Graduate School of Physics and Astronomy. The National Radio Astronomy Observatory is a facility of the National Science Foundation operated under cooperative agreement by Associated Universities, Inc. 
}

%%%%%%%%%%%%%%%%%%%%%%%%%%%%%%%%%%%%%%%%%%
\authorcontributions{S.A.\ Mao, C.\ Heiles, and J.\ West planned the observations; Y.K.\ Ma analysed the data and wrote this article; A.\ Basu developed the \textit{QU}-fitting script.}

%%%%%%%%%%%%%%%%%%%%%%%%%%%%%%%%%%%%%%%%%%
\conflictsofinterest{The authors declare no conflict of interest.} 

\externalbibliography{yes}
\bibliography{ms}

%%%%%%%%%%%%%%%%%%%%%%%%%%%%%%%%%%%%%%%%%%
%% optional
%\sampleavailability{Samples of the compounds ...... are available from the authors.}

%%%%%%%%%%%%%%%%%%%%%%%%%%%%%%%%%%%%%%%%%%
\end{document}